\documentclass[]{iopart}
\usepackage{iopams}
\usepackage{setstack}
\usepackage[]{graphicx}
\usepackage{color}
\usepackage[latin1]{inputenc}

\newcommand{\Si}{\rm Si}
\newcommand{\Cs}{{^{133}\rm Cs}}
\newcommand{\Rb}{{^{87}\rm Rb}}

\begin{document}

\title[]{Does the measured value of the Planck constant depend on the energy of measurements?}
\author{E Massa$^1$, G Mana$^1$\footnote[3]{To whom correspondence should be addressed (g.mana@inrim.it)}, and  M Jentschel$^2$}
\address{$^1$INRIM -- Istituto Nazionale di Ricerca Metrologica, str.\ delle Cacce 91, 10135 Torino, Italy}
\address{$^2$ILL -- Institut Laue-Langevin, 6, rue Jules Horowitz, 38042 Grenoble Cedex 9, France}

\begin{abstract}
The measurement of the Avogadro constant opened the way to a comparison of the watt-balance measurements of the Planck constant with the values calculated from the quotients of the Planck constant and the mass of a particle or an atom. Since the energy scales of these measurements span nine energy decades, these data provide insight into the consistency of our understanding of physics.
\end{abstract}

\submitto{Metrologia}


\section{Introduction}
The Planck constant, $h$, links energy and momentum to the frequency and the wavelength of the wave-function, be it classical or relativistic \cite{Borde}. Its determinations match energy (momentum) and frequency (wavelength) measurements and, according to whether a mechanical, electrical, or thermal system is considered, the measurement result is a value of the $h/m$, $h/e$, or $h/k_B$ ratio, where $m$, $e$, and $k_B$ are a mass, the electron charge, and the Boltzmann constant. Eventually, to determine the Planck constant absolute measurements of mass, charge, or temperature are necessary. The archetype of the electrical determinations is Millikan's photoelectric measurement of the $h/e$ ratio \cite{Millikan}, the modern equivalent of which is the measurement of the Josephson constant by the tunneling of Cooper's pairs in a Josephson junction. As regards the thermodynamic measurements, their archetype is the Planck's black-body determination of the $h/k_B$ ratio \cite{Planck}, the modern equivalent of which is the measurement of the Boltzmann constant by the power of Johnson noise in a resistor.

A number of experiments measured energy or momentum in terms of frequency or wavelength via the Planck and de Broglie equations $E=h\nu$ and $p=h/\lambda$. Energy and momentum are related to mass by the Einstein equation $E=mc^2$ and by $p=mv$, where $v$ is the velocity; therefore, the quotient of the Planck constant and mass is also determined. Since, for atoms and sub-atomic particles, molar masses are well known, these experiments deliver accurate values of the molar Planck constant, $N_A h$, but not of the Planck constant itself. The measurement of the Avogadro constant \cite{NA:PRL,Andreas} opens the way to the estimate of $h$ from the results of these experiments.

The present paper summarizes the results of the $N_A$ determination by counting $^{28}$Si atoms. Next it reviews the determinations of the Planck constant via the $h/(2e)$ determination \cite{Clothier:1989,Funck:1991} and the watt-balance experiments \cite{Kibble:1990,Williams:1998,Steiner:2007,Eichenberger:2011,Robinson:2011,Steiner:2005}. Eventually, it outlines the determinations of the molar Planck constant via the time-of-flight determination of monochromatic neutrons \cite{Krueger:1998,Krueger:1999,Martin:1998,Massa:2009,CODATA:1998}, atom interferometry \cite{CODATA:2002,CODATA:2006,Wicht:2002,Clade:2006a,Clade:2006b}, and atomic and nuclear spectroscopy \cite{CODATA:2006,Dewey:2006,Rainville:2005,Rainville:2004,Thompson:2004}.

Since these experiments rely on different quantum effects the energy scales of which range from less than 1 meV to more than 1 MeV, the ubiquitous presence of the Planck constant give us access to a verification of the measurement capabilities, the understanding of the phenomena underlying the measurements, and the approximation made.

\section{$N_A$ determination}
$N_A$ has been determined by counting the atoms in a mole, exploiting their ordered arrangement in a $^{28}$Si crystal. The crystal and the atom volumes -- $V$ and $a_0^3/8$ -- being measured, the count required calculation of their ratio,
\begin{equation}\label{NA}
 N_A = \frac{8VM(\Si)}{m a_0^3} .
\end{equation}
The determination of the silicon moles, $m/M(\Si)$, required the crystal mass and molar mass -- $m$ and $M(\Si)$ -- to be also measured. The use of a crystal highly enriched with the $^{28}$Si isotope made it possible to determine the molar mass by isotope dilution mass spectroscopy with unprecedented accuracy. A spherical crystal-shape was selected to trace the volume determination back to diameter measurements and to make possible accurate geometrical, chemical, and physical characterizations of the crystal surface. The lattice parameter, $a_0$, and, hence, the atom volume, were measured by combined x-ray and optical interferometry. The results,
\begin{equation}\label{NA-value}
 N_A = 6.02214082(18) \times 10^{23} \;{\rm mol}^{-1} ,
\end{equation}
is the most accurate value so far obtained \cite{NA:PRL,Andreas}.

\section{$h/(2e)$ determination}
When a Josephson device is irradiated with electromagnetic radiation the current-to-voltage relation exhibits steps at quantized voltages. These steps are proportional to the frequency of the irradiating radiation, the proportionality factor being theoretically predicted to be $1/K_J=h/(2e)$, where $K_J$ is the Josephson constant. By combining the results of $K_J$ measurements \cite{Clothier:1989,Funck:1991} with the fine structure constant \cite{CODATA:2006},
\begin{equation}\label{alpha}
 \alpha = \frac{\mu_0 c e^2}{2h} = 7.2973525376(50) \times 10^{-3} ,
\end{equation}
or with the von Kilitzing constant \cite{CODATA:2006},
\begin{equation}\label{RK}
 R_K = \frac{h}{e^2} = 25812.807557(18) \; \Omega ,
\end{equation}
values of $h$ are also obtained \cite{Clothier:1989,Funck:1991}. Since the uncertainty of the $\alpha$ value is negligible, the uncertainty of the $h$ value derived from (\ref{alpha}) is twice that of $K_J$.

\section{$h/m_\mathfrak{K}$ determination}
The technologies required to carry out mechanical determinations of the Planck constant became available only recently \cite{Kibble:1990,Williams:1998,Steiner:2007,Eichenberger:2011,Robinson:2011,Steiner:2005}. The direct way of access to the $h/m_{\mathfrak{K}}$ ratio ($m_{\mathfrak{K}}$ is the mass of the international kilogram prototype) is the watt-balance experiment. This experiment compares virtually the mechanical and electrical powers produced by the motion of a kilogram prototype in the earth gravitational field and by the motion of the supporting coil in a magnetic field. The Planck constant, up to integers corresponding to Josephson and quantum Hall steps, is determined by $m_{\mathfrak{K}}gv \propto h\nu_1\nu_2/4$, where $v$ is the motion velocity, $g$ the acceleration due to gravity, and $\nu_1$ and $\nu_2$ the frequency of the microwaves irradiating a Josephson junction to trace the measurement of the electrical power back to the Josephson and the von Klitzing constants, $K_J=2e/h$ and $R_K=h/e^2$.

\section{$N_A h$ determinations}

\subsection{$h/m({\rm n})$ ratio}
The measurement of the $h/m({\rm n})$ was carried out at the high-flux reactor of the Institut Laue-Langevin in 1998 \cite{Krueger:1998,Krueger:1999}. The $m({\rm n}) v = h/\lambda$ equation was used, by measuring both the wavelength and the velocity -- $\lambda$ and $v$ -- of a monochromatic neutron beam. Monochromaticity was obtained by Bragg-reflection on a silicon crystal, whereas the neutron velocity was determined by time-of-flight measurements. The lattice parameter of a series of different monochromator crystals was determined by comparison against the lattice parameter of a sample of the WASO04 crystal \cite{Krueger:1998,Martin:1998}, the lattice parameter of which has been measured by combined x-ray and optical interferometry \cite{Massa:2009}. The measurement result can be expressed in terms of the molar Planck constant and the neutron molar-mass as
\begin{equation}\label{hmn-value}
 \frac{N_A h}{M({\rm n})} = 3.956033285(287) \times 10^{-7} \; {\rm m}^2{\rm s}^{-1} .
\end{equation}
The neutron molar-mass -- $M({\rm n}) = 1.00866491597(43)$ g/mol -- is determined by comparing its binding energy in the deuterium with the molar-mass difference between the deuterium and hydrogen atoms \cite{CODATA:1998}. The relevant measurement equation is
\begin{equation}
 \big[ M(^1{\rm H}) + M({\rm n}) - M(^2{\rm H}) \big] c^2 = N_A h \nu_D ,
\end{equation}
where the transition frequency $\nu_D$ is measured by nuclear spectroscopy after the capture of a thermal neutron by an hydrogen nucleus. In turn, the $M({\rm n})$ determination required the knowledge of $N_A h$ \cite{CODATA:1998}, which was obtained as will be described in section \ref{hme_ratio}. The vicious circle is avoided by observing that $N_A h \nu/c^2$ is an extremely small correction to the $M(^2{\rm H})-M(^1{\rm H})$ difference. For this reason, the energy scale of this determination is that of thermal neutrons. Since the uncertainty of the $M({\rm n})$ value is negligible, the relative uncertainty of the $N_A h$ value derived from (\ref{hmn-value}) is the same as that of the $h/m({\rm n})$ ratio.

\subsection{$h/m(\Cs)$ and $h/m(\Rb)$ ratios}
The $h/m(\Cs)$ ratio was determined by atom interferometry via the measurement of the recoil frequency-shift $\Delta \nu$ of photons absorbed and emitted by $\Cs$ \cite{CODATA:2002,CODATA:2006,Wicht:2002}. When an atom at rest adsorbs or emits a photon, the photon momentum is balanced by recoil of the atom. Therefore, part of the transition energy is stored in the kinetic energy of the atom and the photon has a higher (adsorption) or lower (emission) frequency than the $\nu_0$ value expected from the energy difference. Eventually, conservation of momentum and energy yields
\begin{equation}
  \frac{m(\Cs) c^2 \Delta \nu}{\nu_0} = h\nu_0
\end{equation}
and, when substituting the molar mass $M(\Cs) = 132.905451932(24)$ g/mol for the $\Cs$ mass \cite{CODATA:2006} and $N_A h$ for the Planck constant, one obtains
\begin{equation}
  \frac{N_A h}{M(\Cs)} = 3.002369432(46) \times 10^{-9} \; {\rm m}^2 {\rm s}^{-1} .
\end{equation}

In the case of the $\Rb$ atom, the recoil velocity $v_{\rm Rb}$ was measured and the conservation of momentum yields $m(\Rb) v_{\rm Rb} = h/\lambda_0$, where $\lambda_0$ is the photon wavelength \cite{CODATA:2006,Clade:2006a,Clade:2006b}. By substituting the molar mass $M(\Rb) = 86.909180526(12)$ g/mol for the $\Rb$ mass \cite{CODATA:2006} and $N_A h$ for the Planck constant, one obtains
\begin{equation}
  \frac{N_A h}{M(\Rb)} = 4.591359287(61) \times 10^{-9} \; {\rm m}^2 {\rm s}^{-1} .
\end{equation}

\subsection{$h/m(e)$ ratio}\label{hme_ratio}
The quotient $h/m(e)$ of the Planck constant and the electron mass can be obtained from the measurement values of the Rydberg and the fine-structure constants. The Rydberg constant, $R_\infty = 10973731.568527(73)$ m$^{-1}$, is determined by comparing the frequencies of the photons emitted in transitions of hydrogen and deuterium to the theoretical expression for the atom energy. Since $R_\infty$ is given by
\begin{equation}\label{Rydberg}
 R_\infty = \frac{\alpha^2 m(e) c}{2h} ,
\end{equation}
where $\alpha = 7.2973525376(50) \times 10^{-3}$ \cite{CODATA:2006}, by writing the electron mass $m(e)$ in terms of molar mass, $M(e) = N_A m(e)$, where $M(e)=5.4857990943(23) \times 10^{-4}$ g/mol \cite{CODATA:2006}, one obtains
\begin{equation}\label{hme-value}
 \frac{N_A h}{M(e)} = \frac{\alpha^2 c}{2R_\infty} = 7.27389504(10) \times 10^{-4} \; {\rm m}^2{\rm s}^{-1} .
\end{equation}

\subsection{neutron binding energy}
The molar Planck constant was determined by measuring the frequencies of the $\gamma$ photons emitted in the cascades from the neutron capture state to the ground state in the reaction n + $^n$X $\rightarrow$ $^{n+1}$X$^*$ $\rightarrow$ $^{n+1}$X + $\gamma$. Frequencies were determined at the Institut Laue-Langevin in terms of the lattice parameter of a diffracting crystal traced back to optical wavelengths by combined x-ray and optical interferometry \cite{Dewey:2006}. The $N_A h$ measurement is based on the comparison of the total energy of the emitted $\gamma$ rays and the mass defect $\Delta m$ between the initial and final states; energy and mass are compared by the Einstein's and Planck's relations $E=\Delta m c^2$ and $E=h\nu$ \cite{Rainville:2005}. The mass difference was measured by simultaneous comparisons of the cyclotron frequencies of ions of the initial and final isotopes confined in a Penning trap \cite{Rainville:2004,Thompson:2004}. By combining
\numparts\begin{equation}
 \big[M(^n{\rm X}) + M({\rm n}) - M(^{n+1}{\rm X}) \big] c^2 = N_A h \nu_{n+1}
\end{equation}
and
\begin{equation}
 \big[M(^1{\rm H}) + M({\rm n}) - M(^2{\rm H}) \big] c^2 = N_A h \nu_D ,
\end{equation}
the comparison can be expressed as
\begin{equation}\fl
 \big[M(^n{\rm X}) + M(^2{\rm H}) - M(^{n+1}{\rm X}) - M(^1{\rm H}) \big] c^2 = N_A h(\nu_{n+1} - \nu_D) .
\end{equation}\endnumparts
The molar Planck constant values obtained are
\numparts\begin{equation}\label{NAh-1}
 N_A h = 3.956033285(287) \times 10^{-7} \; {\rm J\, s\, mol}^{-1}
\end{equation}
from the n + $^{28}$Si $\rightarrow$ $^{29}$Si + $\gamma$ reaction and
\begin{equation}\label{NAh-2}
 N_A h = 3.956033285(287) \times 10^{-7} \; {\rm J\, s\, mol}^{-1}
\end{equation}\endnumparts
from the n + $^{32}$S $\rightarrow$  $^{33}$S reaction.

\begin{table}[b]
\caption{Values of the Planck constant. The values calculated from quotient of $h$ and the neutron mass, the mass of a particle or an atom, and the neutron binding energy depend on the same measured value of $N_A$ \cite{Andreas}.}
\begin{center}
\begin{tabular}{lllr}
\hline\hline
method &energy / eV &$10^{34} h$ / Js  &reference\\
\hline\hline
\multicolumn{3}{c}{quotient of $h$ and the electron charge} \\
 $h/(2e)$  (NMI 89)                  &$10^{-3}$ &$6.6260684(36) $  &\cite{Clothier:1989} \\
 $h/(2e)$  (PTB 91)                  &$10^{-3}$ &$6.6260670(42) $  &\cite{Funck:1991} \\
\multicolumn{3}{c}{watt-balance experiments} \\
 $h/m_{\mathfrak{K}}$ (NPL 1990)     &$10^{-3}$ &$6.6260682(13) $  &\cite{Kibble:1990} \\
 $h/m_{\mathfrak{K}}$ (NIST 1998 )   &$10^{-3}$ &$6.62606891(58)$  &\cite{Williams:1998} \\
 $h/m_{\mathfrak{K}}$ (NIST 2007)    &$10^{-3}$ &$6.62606901(34)$  &\cite{Steiner:2007} \\
 $h/m_{\mathfrak{K}}$ (METAS 2011)   &$10^{-3}$ &$6.6260691(20) $  &\cite{Eichenberger:2011} \\
 $h/m_{\mathfrak{K}}$ (NPL 2010)     &$10^{-3}$ &$6.6260712(13) $  &\cite{Robinson:2011} \\
\multicolumn{3}{c}{weighted mean (1 meV)} \\
                                     &$10^{-3}$ &$6.62606904(28)$ \\
\hline
\multicolumn{3}{c}{quotient of $h$ and the neutron mass} \\
 $N_A h/M({\rm n})$                  &$10^{-2}$ &$6.62606887(52)$ &\cite{Krueger:1999,Andreas} \\
\hline
\multicolumn{3}{c}{quotient of $h$ and the mass of a particle or of an atom} \\
 $N_A h/M(e)$                        &1 &$6.62607003(20) $ &\cite{CODATA:2006,Andreas} \\
 $N_A h/M({\rm Cs})$                 &1 &$6.62607000(22) $ &\cite{CODATA:2006,Andreas} \\
 $N_A h/M({\rm Rb})$                 &1 &$6.62607011(22) $ &\cite{CODATA:2006,Andreas} \\
\multicolumn{3}{c}{weighted mean (1 eV)} \\
                                     &1 &$6.62607005(21) $ \\
\hline
\multicolumn{3}{c}{quotient of $h$ and the neutron binding energy} \\
 $h/\Delta m(^{29}{\rm Si})$  &$10^6$ &$6.6260764(53) $ &\cite{Rainville:2005,Andreas} \\
 $h/\Delta m(^{33}{\rm S})$   &$10^6$ &$6.6260686(34) $ &\cite{Rainville:2005,Andreas} \\
\multicolumn{3}{c}{weighted mean (1 MeV)} \\
                                     &$10^6$ &$6.6260709(29) $ \\
\hline\hline
\end{tabular} \label{planck-values-table} \end{center} \end{table}

\begin{figure}
\centering
\includegraphics[width=75mm]{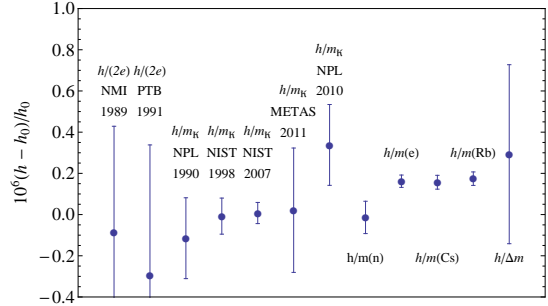}
\caption{Comparison of the $h$ determinations given in table \ref{planck-values-table}. The reference is the CODATA 2006 value, $h_0=6.62606896\times 10^{-34}$ Js \cite{CODATA:2006}.}\label{planck-values-plot}
\end{figure}

\begin{figure}[b]
\centering
\includegraphics[width=60mm]{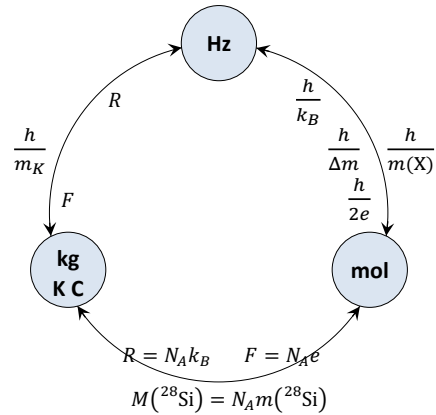}
\caption{Interconnections between the $N_A$-related units. The arrows indicate the interlinked units. The kg, K, and C units represent macroscopic (molar) quantities. The mol unit represents the corresponding microscopic quantity.}\label{triangle}
\end{figure}

\section{Conclusions}
When atoms and sub-atomic particles are used, the $h/m$ quotient does not imply a determination of $h$, but of the $N_A h$ product. The $N_A$ value allows $h$ values to be obtained; they are summarized and compared with the values obtained from the $h/(2e)$ and $h/m_\mathfrak{K}$ measurements in table \ref{planck-values-table} and Fig.\ \ref{planck-values-plot}. This comparison is a test of the measurement capabilities and of the understanding of the underlying phenomena; the interconnections between the mass-related units are shown in Fig.\ \ref{triangle}. Given a value of the Planck constant, frequency measurements are related to both microscopic and macroscopic mass measurements -- represented by the mole and the kilogram.

In Fig.\ \ref{triangle}, the measured values of $h/(2e)$ and $h/m_\mathfrak{K}$ depend on solid-state physics through the Josephson and von Klitzing constants. The $h/m({\rm n})$ value depends on the wave-particle duality; those of the $h/m(e)$, $h/m(\Cs)$, and $h/m(\Rb)$ ratios depend on atomic physics. Eventually, measurement of the neutron binding energy in $^{29}$Si and $^{33}$S depends on the mass-energy-frequency equivalence and nuclear physics. The determination of $N_A$ by counting $^{28}$Si atoms links these measurements and tests the result consistency.

In table \ref{planck-values-table} and Fig.\ \ref{planck-values-plot}, the values calculated from ratio of $h$ and the neutron mass, the mass of a particle or of an atom, and the neutron binding energy have been obtained by adoption of the same $N_A$ value; their scatter reflects the variations in the fine-structure constant values that are inferred from these quotients \cite{CODATA:2006}. The uncertainty of the $h$ values derived from $h/m(e)$, $h/m(\Cs)$, and $h/m(\Rb)$ is governed by the $N_A$ uncertainty; therefore, they are strongly correlated. However, the $N_A$ uncertainty affects only marginally the $h$ values derived from $h/m({\rm n})$ and $h/\Delta m$. Though the most accurate $h$ values -- the NIST 2007 one and that derived by combining the $h/m(e)$ and $N_A$ values -- are mutually inconsistent, none is totally out of scale.

\begin{figure}
\centering
\includegraphics[width=75mm]{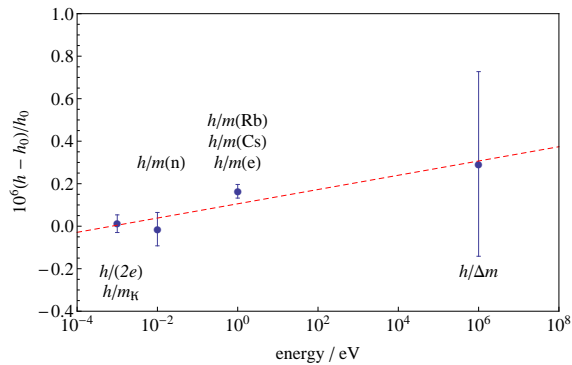}
\caption{Planck constant values given in table \ref{planck-values-table} sorted according the energy of the underlying measurement principle. The values obtained with experiments carried out at the same energy have been averaged. The reference is the CODATA 2006 value, $h_0=6.62606896\times 10^{-34}$ Js \cite{CODATA:2006}.}\label{energy}
\end{figure}

As shown in Fig.\ \ref{energy}, the energies of the phenomena implicated in these measurements span from less than 1 meV to more than 1 MeV; from the comparison of the measured $h$ values, the relevant models appear consistent to within a relative accuracy of $1\times 10^{-7}$. The figure shows a slight increase of the measured $h$ values with increasing measurement energies: the value derived from the $h/m({\rm n})$ ratio agrees with the best watt-balance determination (the NIST 2007 one); the values relying on atomic and nuclear spectroscopy are slightly greater. Since inconsistencies could also be a clue for incomplete understanding of the implicated phenomena, this motivates efforts to reduce the measurement uncertainties and to increase the confidence in the measured values.

\ack
This research received funding from the European Community's Seventh Framework Programme, ERA-NET Plus, under the iMERA-Plus Project - Grant Agreement No. 217257.

\end{document}